\documentclass[superscriptaddress,aps,preprintnumbers,amsmath,amssymb,prd,nofootinbib,preprint]{revtex4-1}
\pdfoutput=1

\usepackage{graphicx}
\usepackage{epstopdf}
\usepackage{dcolumn}
\usepackage{bm}
\usepackage{hyperref}
\usepackage{color}
\usepackage{amsmath,amssymb}
\usepackage[sort&compress]{natbib}
\usepackage{float}
\usepackage{mathtools}

\numberwithin{equation}{section}

\makeatletter
\renewcommand{\p@subsection}{}
\renewcommand{\p@subsubsection}{}
\makeatother

\begin{document}


\title{
Increasing Temperature toward the Completion of Reheating \\
}
\preprint{LCTP-20-15}

\author{Raymond T. Co}
\affiliation{Leinweber Center for Theoretical Physics, Department of Physics, University of Michigan, Ann Arbor, MI 48109, USA}
\author{Eric Gonzalez}
\affiliation{Leinweber Center for Theoretical Physics, Department of Physics, University of Michigan, Ann Arbor, MI 48109, USA}
\author{Keisuke Harigaya}
\affiliation{School of Natural Sciences, Institute for Advanced Study, Princeton, NJ 08540, USA}

\begin{abstract}
Reheating is a process where the energy density of a dominant component of the universe other than radiation, such as a matter component, is transferred into radiation. It is usually assumed that the temperature of the universe decreases due to cosmic expansion even during the reheating process, in which case the maximal temperature of the universe is much higher than the reheat temperature. We point out that the temperature of the universe during reheating may in fact increase in well-motivated scenarios. We derive the necessary conditions for the temperature to increase during reheating and discuss concrete examples involving a scalar field. We comment on implications for particle physics and cosmology due to an increasing temperature during reheating.
\end{abstract}

\date{\today}

\maketitle

\section{Introduction}
\label{sec:intro}

The observable universe is mysteriously homogeneous and isotropic even though different patches of the universe are causally disconnected. This is known as the horizon problem. The curvature of the universe is flat to a very high precision, which requires an incredible amount of tuning if one attempts to offer an explanation by the initial condition. This is referred to as the flatness problem. Both of these problems are elegantly solved by an early epoch of cosmic inflation~\cite{Guth:1980zm} (see also~\cite{Kazanas:1980tx}), where the universe expands by many orders of magnitude due to a positive potential energy. This large potential energy will eventually create the radiation constituent of the present universe through the process called reheating~\cite{Kolb:1990vq}. The universe may also be reheated again due to long-lived particles or fields. Specifically, string theory and models of supersymmetry generically contain flat directions in the scalar potential that lead to light fields. The light fields tend to overclose the universe or cause cosmological problems unless they are thermalized sufficiently early. Thermalization of light fields reheats the universe again if their energy density dominates. 

\begin{figure}
	\includegraphics[width= \linewidth]{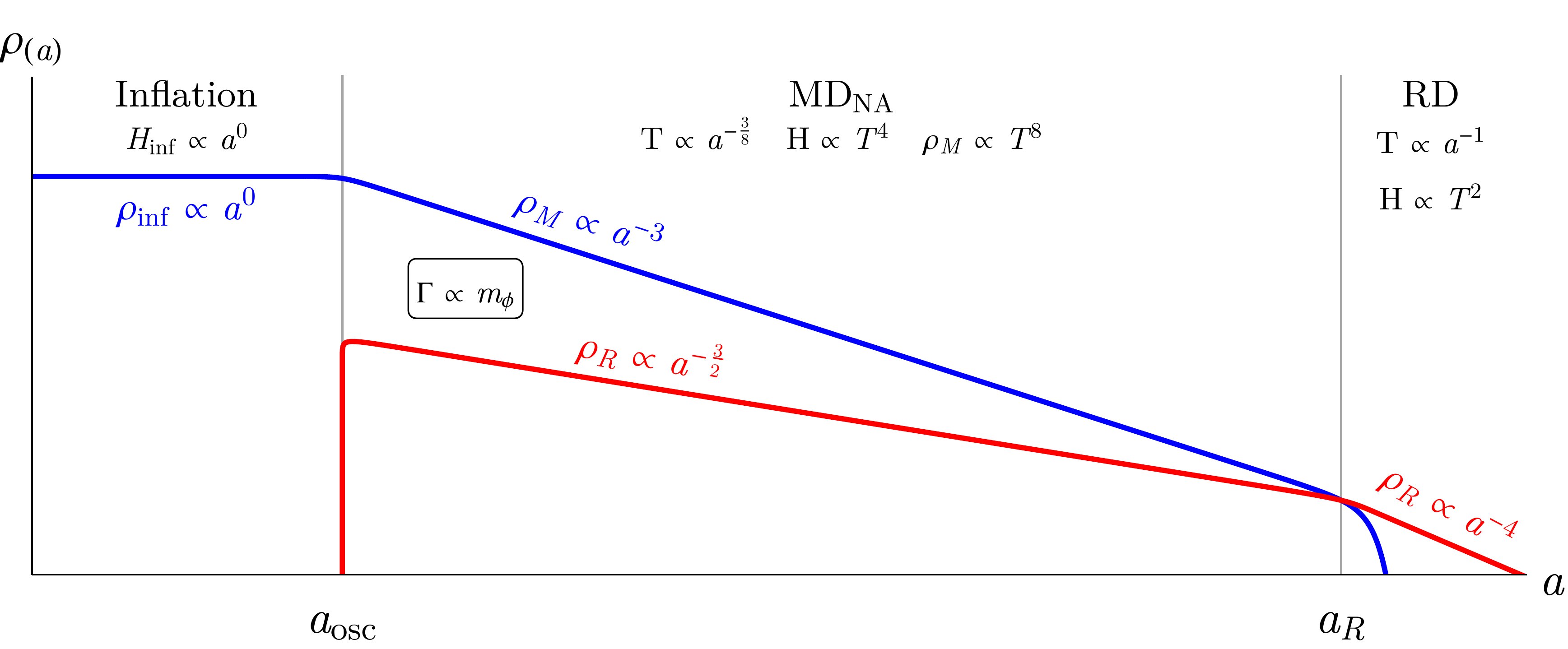}
	\caption{Evolution of the energy densities of the inflaton $\rho_M$ and of radiation $\rho_R$ with the scale factor $a$ on logarithmic scales, in the standard case of the perturbative decay of matter into radiation for inflationary reheating. When the inflaton starts to oscillate at $a_{\rm osc}$ immediately after inflation, the inflaton decay brings radiation to its maximum value. Radiation then scales as $\rho_R\propto a^{-3/2}$ during the non-adiabatic matter dominated (MD$_{\rm NA}$) era. When decays are efficient, reheating ends at $a_R$ and a radiation dominated (RD) era ensues. The scaling of the temperature $T$ and Hubble parameter $H$ is also labeled. This cosmology arises when the rate $\Gamma$ is constant in time.}
	\label{fig:pert_decay}
\end{figure}

The conventional scenario for reheating assumes the perturbative decays of matter into radiation. For inflationary reheating, as illustrated in Fig.~\ref{fig:pert_decay}, the radiation energy density of the universe increases in a short period of time immediately after inflation to a maximum value and decreases for the remaining duration of reheating due to cosmic expansion~\cite{Kolb:1990vq}. Radiation created from reheating will reach thermal equilibrium and can be characterized by a temperature. The reheating period ends at the so-called reheat temperature $T_R$, where matter and radiation have comparable energy densities and the dissipation rate is also comparable to the Hubble expansion rate. For reheating by generic light fields, despite the term {\it reheating}, the temperature of the universe never increases but simply decreases less quickly than without this injection of energy. In this paper, we point out that, if the dissipation rates of the inflaton and light fields increase with time, the temperature may increase or remain constant throughout reheating. We provide examples where the dynamics of the scalar fields itself leads to the increasing dissipation rate.

We point out several implications of such reheating eras where the universe does not cool down during reheating as in the conventional scenario. In theories requiring a high reheat temperature such as thermal leptogenesis~\cite{Fukugita:1986hr,Giudice:2003jh,Buchmuller:2004nz}, conventional reheating predicts that $T_{\rm max}$ is typically much larger than $T_R$~\cite{Harigaya:2013vwa,Mukaida:2015ria} and a large $T_{\rm max}$ may restore symmetries such as the Peccei-Quinn~\cite{Peccei:1977hh,Peccei:1977ur,Kim:1979if,Shifman:1979if,Zhitnitsky:1980tq,Dine:1981rt}, left-right~\cite{Pati:1974yy, Mohapatra:1974hk,Senjanovic:1975rk,Beg:1978mt,Mohapatra:1978fy,Babu:1988mw,Babu:1989rb,Albaid:2015axa,Hall:2018let,Dunsky:2019api,Hall:2019qwx}, or CP symmetries~\cite{Lee:1973iz,Nelson:1983zb,Barr:1984qx,Bento:1991ez,Dine:2015jga} and cause cosmological domain wall problems~\cite{Zeldovich:1974uw,Sikivie:1982qv}. If the temperature instead increases or remains constant during reheating, the maximum temperature $T_{\rm max}$ achieved is only as high as the reheat temperature $T_R$ and symmetry restoration is prevented. On the other hand, if the critical temperature of a phase transition is below $T_R$, the universe will undergo the phase transition twice as the universe is heated up and cools back down. It is known that strong first order phase transitions can generate observable gravitational waves~\cite{Witten:1984rs}. An increasing temperature may also lead to additional signals because of the extra phase transition. Lastly, the understanding of dark matter production during the early stage of reheating~\cite{Chung:1998rq,Giudice:2000ex,Harigaya:2014waa,Garcia:2017tuj,Garcia:2018wtq,Harigaya:2019tzu, Garcia:2020eof} will also be significantly modified by a qualitatively different temperature and time relation.

In Sec.~\ref{sec:reheat}, we analytically derive the cosmological evolution of the matter and radiation energy densities and distinguish different reheating scenarios depending on the functional form of the dissipation rate. In Sec.~\ref{sec:models}, we discuss various scenarios where the novel reheating eras arise.
In Sec.~\ref{sec:discussions}, we summarize and elaborate on the potential impact on particle physics.

\section{Cosmological Evolution for General Reheating}
\label{sec:reheat}

In this work, we point out the possibility of cosmological eras where the temperature of the radiation bath \emph{increases} with time. To explore the condition for such a period, we consider a generic time-dependent rate $\Gamma$ for dissipation of matter with the energy density $\rho_M$ into the radiation bath with the energy density $\rho_R$. The Boltzmann equations read\footnote{Here the motion of the matter field is assumed to be driven coherently by the potential. The term associated with thermal fluctuations in the equation of motion is negligible and thus omitted.}
\begin{align}
\label{eq:boltzmann-matter}
\dot{\rho}_M + 3H\rho_M & = -\Gamma\rho_M \\
\label{eq:boltzmann-radiation}
\dot{\rho}_R +4H\rho_R & = \Gamma\rho_M
\end{align}
for matter and radiation respectively, where $H = \sqrt{\rho_M+\rho_R}/\sqrt{3} M_{\rm Pl}$ is the Hubble expansion rate and $M_{\rm Pl}$ is the reduced Planck mass. We generalize the dissipation rate
\begin{equation}
\label{eq:Gamma_n_k}
\Gamma = b T^n \left(\frac{a}{a_i}\right)^k,
\end{equation}
where $T$ is the temperature, $b$ is a model-dependent coupling, and $a$ and $a_i$ are the scale factor and initial scale factor respectively. The physical origins of the scale factor dependence are provided in Sec.~\ref{sec:models}. Throughout this paper, we make the assumption that thermalization is efficient and radiation sourced by dissipation of matter instantaneously reaches thermal equilibrium, allowing us to write $\rho_R = \pi^2 g_* T^4 / 30$ at all times, where $g_*$ is the effective degrees of freedom in the thermal bath. In this case, the rate can be expressed in terms of the radiation energy density and $\beta \equiv b \left( 30 / \pi^2 g_* \right)^{n/4}$
\begin{equation}
\label{eq:Gamma_rhoR_a}
\Gamma = \beta\ \rho_R^{n/4} \left(\frac{a}{a_i}\right)^k .
\end{equation}

\begin{figure}
	\includegraphics[width= 0.6\linewidth]{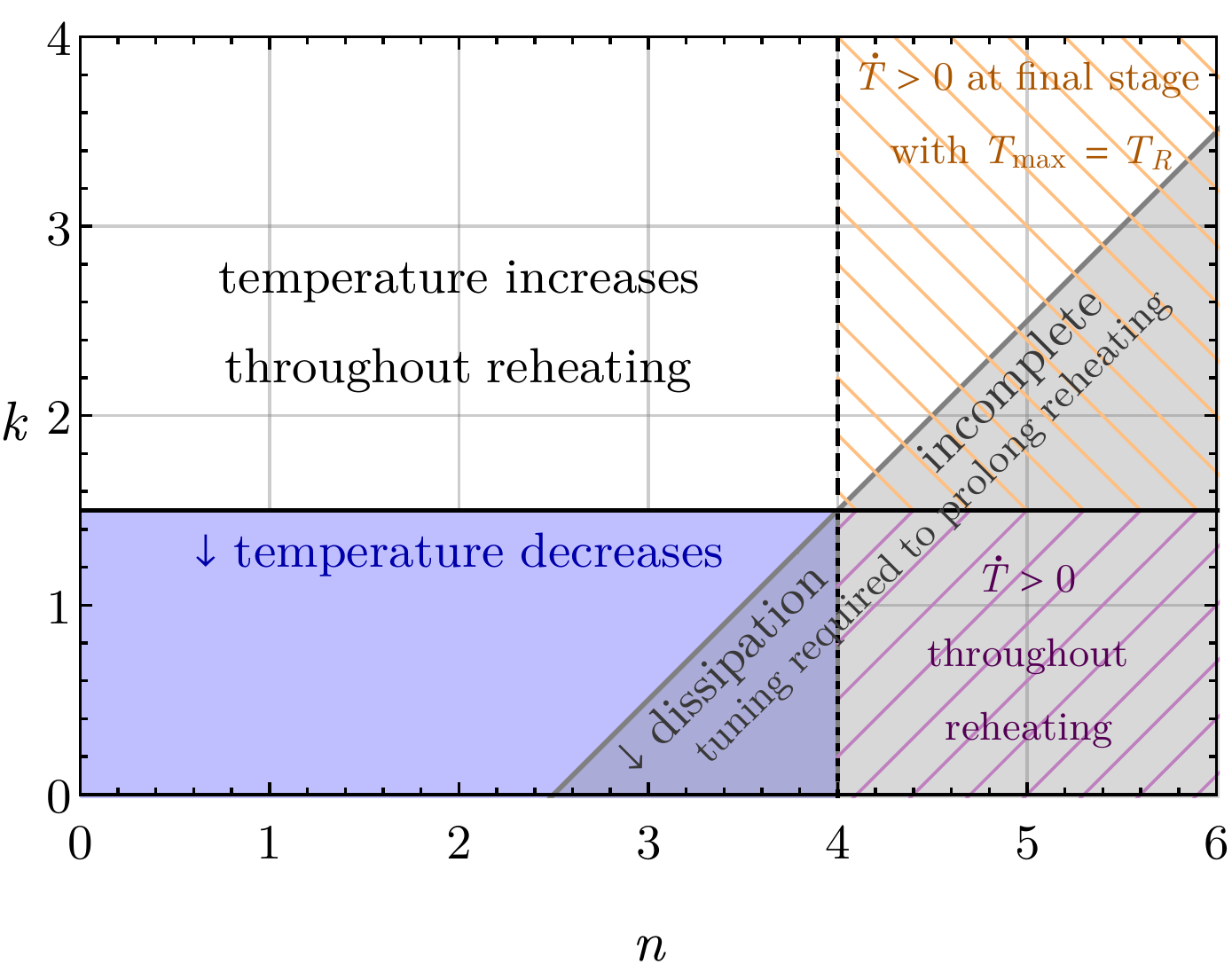}
	\caption{Different reheating scenarios as a result of different dissipation rates. In our convention described in Sec.~\ref{sec:reheat}, the scenarios are distinguished according $n$ and $k$, which parameterize the dependence of the rate $\Gamma \propto T^n a^k$ on the temperature $T$ and the scale factor $a$. This diagram excludes instantaneous reheating, which is possible but not of our interest in this paper.}
	\label{fig:phase-diagram}	
\end{figure}

We study this dissipation rate in two settings: inflationary reheating and generic reheating that comes after early radiation and matter domination. We note that inflation exponentially dilutes the pre-existing radiation. Still, some radiation is created immediately after inflation by the perturbative decays of the inflaton and/or other scalar fields in a way similar to the sudden increase of $\rho_R$ shown in Fig.~\ref{fig:pert_decay}, providing the thermal bath for the inflaton to interact with for $n\neq 0$. In the generic reheating setting, the initial radiation bath comes from the redshifted radiation in the early radiation dominated era. When analyzing the effects of this dissipation rate, we find that the initial radiation bath can play a critical role in determining the final dissipation condition. The initial radiation bath is important for cosmologies where $n\geq4$, and we will detail these solutions in the following subsections. As a starting point, however, we restrict ourselves to scenarios where $n<4$ and the contribution to radiation from matter dissipation dominates the initial radiation. In this case, with Eq.~(\ref{eq:boltzmann-radiation}), we can estimate the amount of radiation produced per Hubble time by 
\begin{equation}
\label{eq:naive_estimate}
\rho_R \simeq \frac{\Gamma}{H} \rho_M .
\end{equation}
Upon using Eq.~(\ref{eq:Gamma_rhoR_a}) and assuming dissipation is not yet efficient in depleting matter, i.e.~$\rho_M \propto a^{-3}$, we obtain the radiation energy density as a function of the scale factor
\begin{equation}
\label{eq:rhoR_T_est}
    \rho_R \simeq \left( \beta\frac{\rho_{M,i}}{H_i} \left(\frac{a}{a_i}\right)^{k-\frac{3}{2}}\right)^{\scalebox{1.01}{$\frac{4}{4-n}$}},
    \hspace{1cm}
    T \propto \left(\frac{a}{a_i}\right)^{\scalebox{1.01}{$\frac{2k-3}{2(4-n)}$}},
\end{equation}
where $\rho_{M,i}$ and $H_i$ are the initial matter energy density and the initial Hubble parameter respectively at the time when the dissipation products dominate over the initial radiation bath. Fig.~\ref{fig:phase-diagram} describes the temperature scaling and reheating condition of the $n$-$k$ parameter space. Our result in Eq.~(\ref{eq:rhoR_T_est}) is valid for $n<4$ in the diagram. When $n<4$ and $k>3/2$, the radiation energy density grows with the scale factor as the universe evolves forward in time as indicated by Eq.~(\ref{eq:rhoR_T_est}). This implies that the temperature increases while dissipation of matter is active, and that the reheat temperature $T_R$ is the maximum temperature reached throughout the evolution. The white region in Fig.~\ref{fig:phase-diagram} corresponds to the scenario with $n<4$ and $k>3/2$. In cosmologies where $n<4$ and $k<3/2$, we find no period of increasing temperature and that the evolution is qualitatively the same as that of perturbative decays ($n = 0$ and $k = 0$). This corresponds to the blue shaded region in Fig.~\ref{fig:phase-diagram}. When $k=3/2$, as on the black solid line of Fig.~\ref{fig:phase-diagram}, the temperature will remain constant, and the value can be obtained from Eq.~(\ref{eq:naive_estimate}) as long as dissipation is the dominant source of radiation.

We provide both the analytic treatment and the numerical evaluation in the following subsections for all $n$ and $k$. Here we describe the final results presented in Fig.~\ref{fig:phase-diagram}. The rigorous analysis for $n<4$ confirms our earlier results in the white and blue shaded regions. In the gray region, we find that matter fails to transfer all its energy to radiation; hence, while reheating can occur, dissipation does not complete. In this case, if the dissipation rate is tuned close to the Hubble rate at the end of inflation $H_{\rm inf}$, a long reheating epoch exists but complete dissipation of matter is still absent. On the other hand, if $\Gamma \lessgtr H_{\rm inf}$, then reheating does not occur or instantaneously occurs, respectively. Furthermore, in the purple hatched region, the temperature monotonically increases during reheating. The orange hatched region corresponds to an intriguing case where the temperature only and abruptly increases near the end of reheating, whether or not dissipation completes.

\subsection{Exact Solutions of the Boltzmann Equations}
Our previous estimate in Eq.~(\ref{eq:naive_estimate}) made the strong assumption that the initial component of radiation is negligible. A full analysis of Eqs.~(\ref{eq:boltzmann-matter}) and (\ref{eq:boltzmann-radiation}) will reveal that the estimate is not valid when $n\geq4$, as the initial radiation bath will significantly affect reheating dynamics. Rewriting Eqs.~(\ref{eq:boltzmann-matter}) and (\ref{eq:boltzmann-radiation}) in terms of comoving energy densities, $X_M \equiv \rho_M (a/a_i)^3$ and $X_R \equiv \rho_R (a/a_i)^4$, we obtain
\begin{align}
\dot{X}_M = -\Gamma X_M \\
\dot{X}_R = \Gamma X_M \left( \frac{a}{a_i} \right). 
\end{align}
At early times when the rate is negligible $\Gamma t \ll 1$, the solution reads $X_M(t) = X_M(0)$. Although the rate is negligible with respect to depleting the matter density, it can still be significant in providing the dominant source of radiation. 

The Boltzmann equation for the comoving radiation energy density is
\begin{equation}
\dot{X}_R = \Gamma X_M \left( \frac{a}{a_i} \right) = \beta\ X_R^{n/4} X_M \left(\frac{a}{a_i}\right)^{k-n+1},
\end{equation}
which one can solve in terms of the scale factor
\begin{align}
\int_{X_{R,i}}^{X_R} \frac{dX_R}{X_R^{n/4}} & = \beta\ \frac{X_{M}}{a_i H_i} \int^{a}_{a_i} \left(\frac{a}{a_i}\right)^{k-n+\frac{3}{2}} da \\
\label{eq:XR_sol}
\int_{X_{R,i}}^{X_R} \frac{dX_R}{X_R^{n/4}} & = \beta\ \frac{X_{M}}{H_i} \frac{1}{k-n+\frac{5}{2}} \left(\left(\frac{a}{a_i}\right)^{k-n+\frac{5}{2}}-1\right).
\end{align}
We discuss different branches of solutions in the following subsections and summarize the qualitative results in Fig.~\ref{fig:phase-diagram}.

\subsection{Solution for $n<4$}
\label{subsec:small_n}
For the case of $n<4$, the solution of Eq.~(\ref{eq:XR_sol}) is of the form
\begin{equation}
\label{eq:full_expression}
X_R^{(4-n)/4} = X_{R,i}^{(4-n)/4} + \beta \frac{X_{M}}{H_i} \frac{4-n}{4 \left(k-n+\frac{5}{2} \right)} \left(\left(\frac{a}{a_i}\right)^{k-n+\frac{5}{2}}-1\right).
\end{equation}

\begin{figure}
	\includegraphics[width= \linewidth]{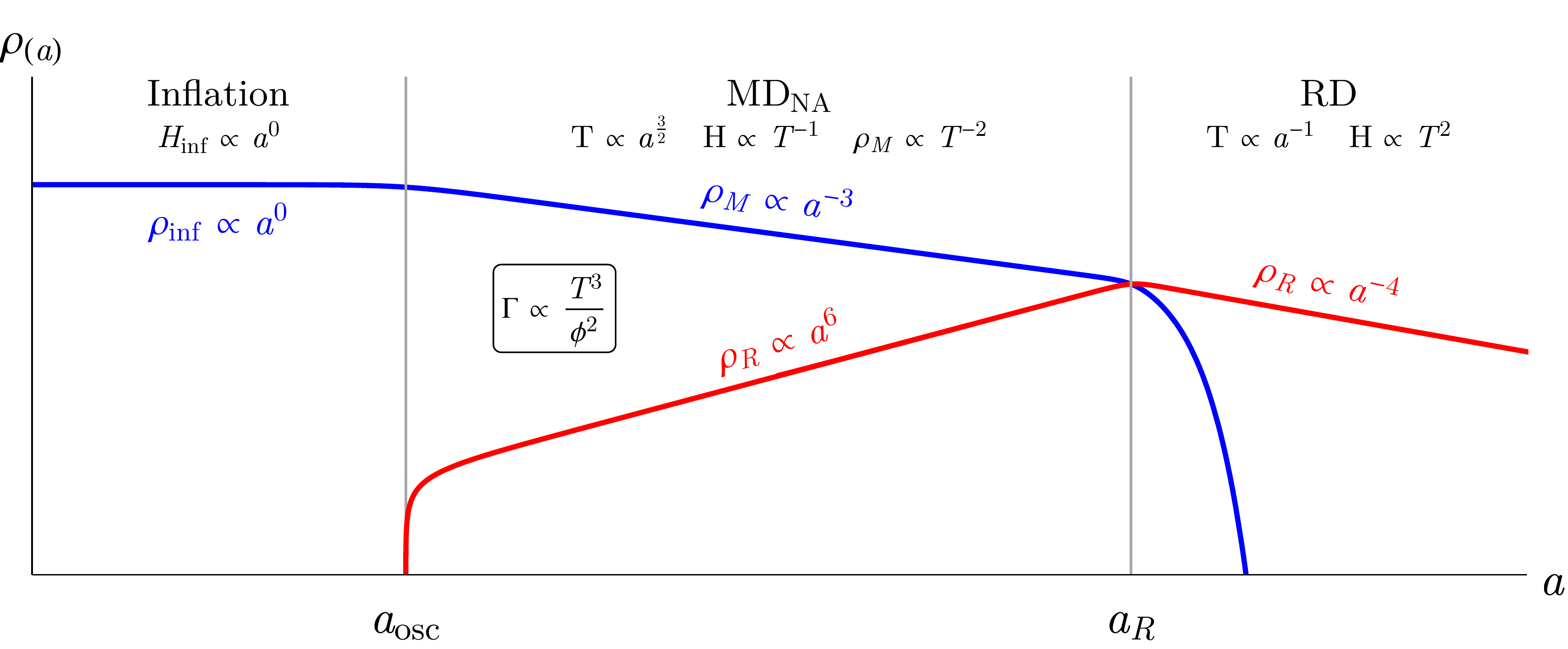}
	\includegraphics[width= \linewidth]{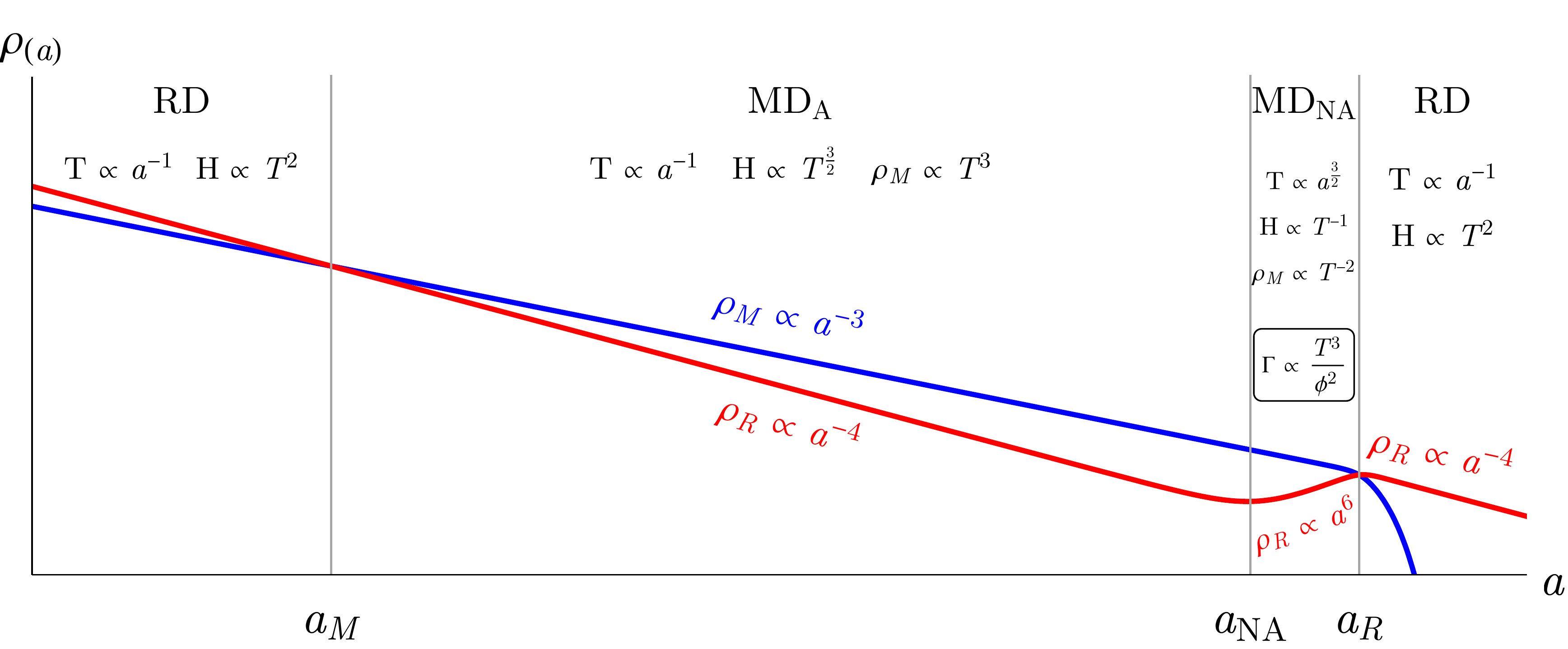}
	\caption{Evolution of the matter energy density $\rho_M$ and radiation energy density $\rho_R$ as a function of the scale factor $a$ with dissipation parameters $n=3$ and $k=3$. A concrete example in Sec.~\ref{sec:models} with a rotating or fluctuating scalar field $\phi$ realizes the dissipation rate $\Gamma \propto T^3 \phi^{-2} \propto T^3 a^3$ (assuming a quadratic potential). The upper panel shows reheating after inflation. An initial radiation bath is created from perturbative decays of matter immediately after inflation ends at $a_{\rm osc}$, similar to Fig.~\ref{fig:pert_decay}, and temperature-dependent processes dominate afterwards. We obtain a cosmology with a period of increasing temperature prior to reheating at $a_R$. This solution shows that reheating is achieved at the maximum temperature $T_{\rm max}$ and is preceded by a period of increasing temperature where $\rho_R\propto a^6$.
	The lower panel is a numerical solution which begins during an early radiation dominated era, followed by a matter dominated era at $a_M$ with adiabatic expansion (MD$_{\rm A}$). Once dissipation has a sizable effect, we enter a non-adiabatic matter dominated era (MD$_{\rm NA}$) at $a_{\rm NA}$ and the temperature increases until reheating completes at $a_R$.}
	\label{fig:n3k3}	
\end{figure}

\begin{figure}
	\includegraphics[width= \linewidth]{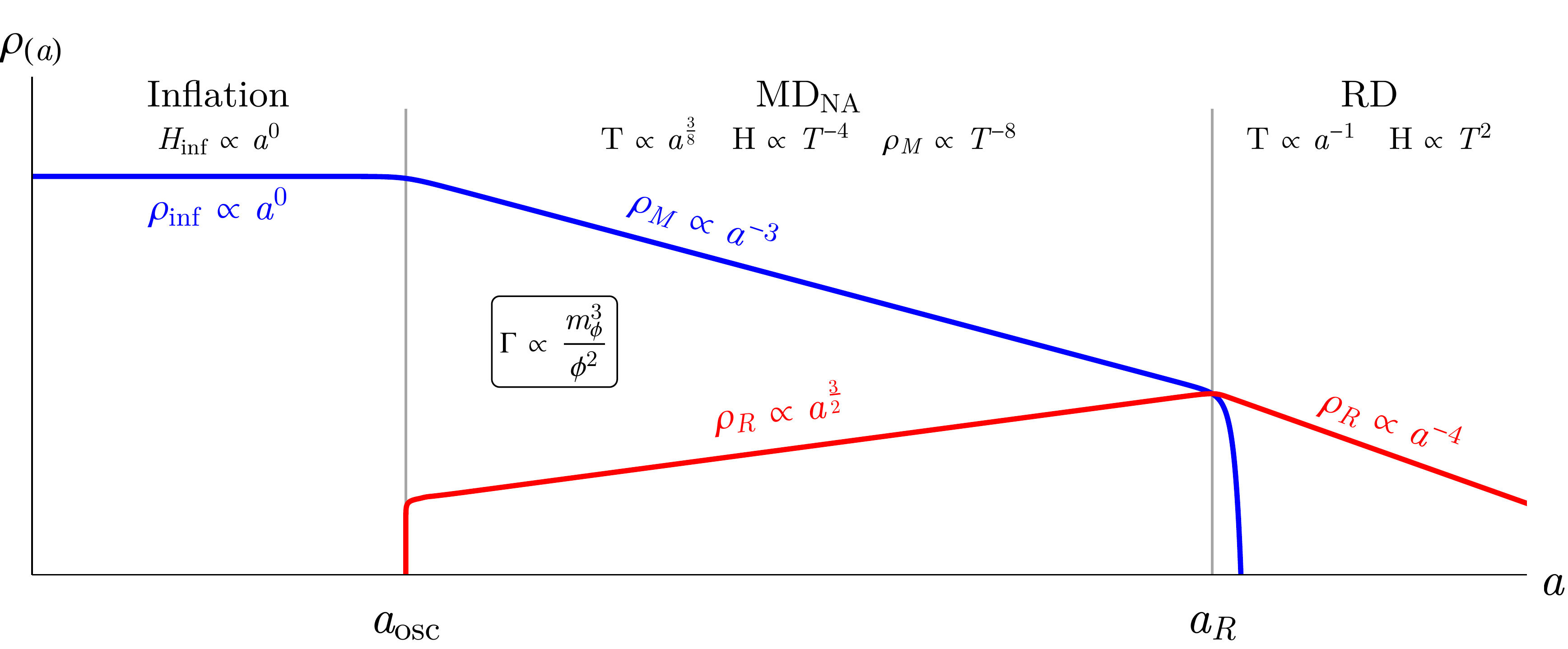}
	\includegraphics[width= \linewidth]{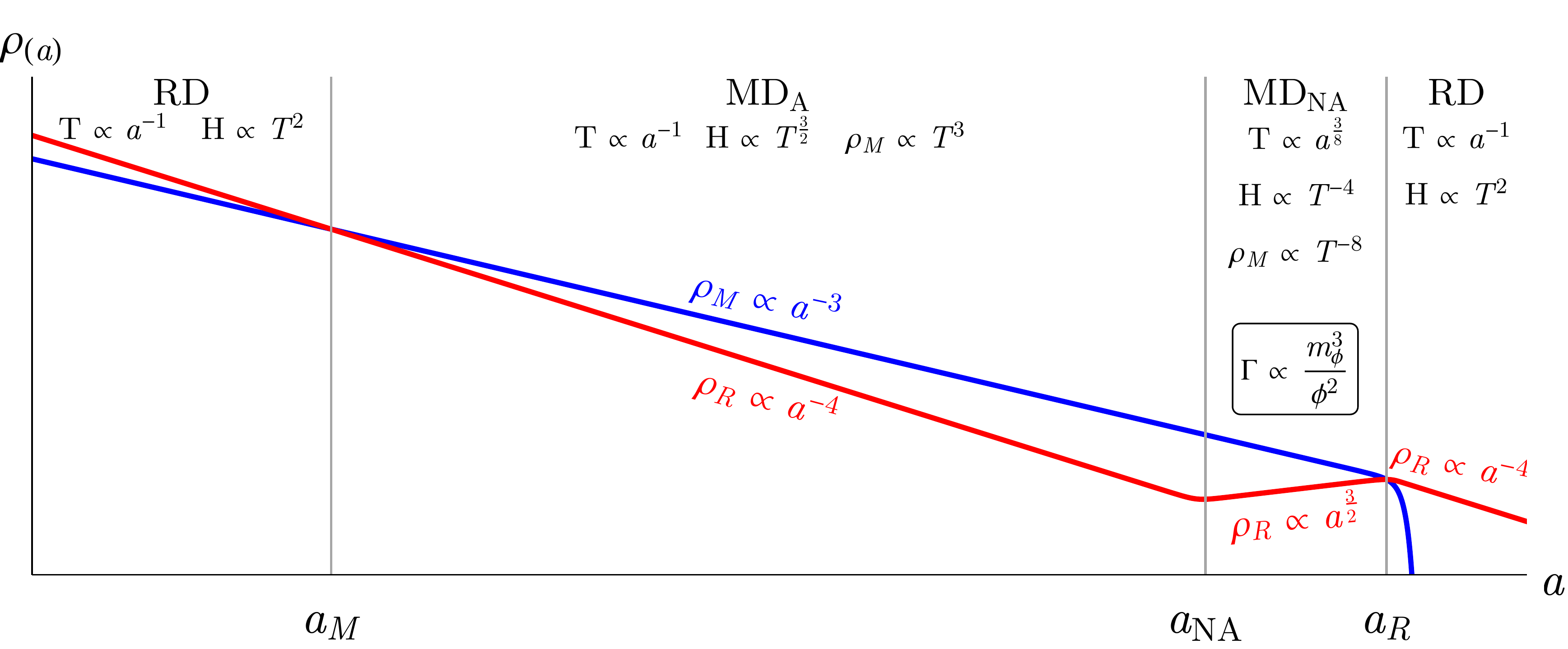}
	\caption{Evolution of the matter energy density $\rho_M$ and radiation energy density $\rho_R$ as a function of the scale factor $a$ with dissipation parameters $n=0$ and $k=3$. In typical models of large field inflation, a large inflaton mass and low reheat temperature may be of interest. In such cases, the thermal effect is negligible and the rate would then depend on the inflaton mass $\Gamma \propto m_\phi^3/\phi^2$. Qualitatively similar to the upper panel of Fig.~\ref{fig:n3k3}, radiation scales as $\rho_R\propto a^{3/2}$ and the temperature increases at a rate $T\propto a^{3/8}$. Reheating completes when $\Gamma \simeq H$ at $a_R$. The lower panel shows an early RD era followed by an adiabatic MD era at $a_M$ and a non-adiabatic MD era at $a_{\rm NA}$. The temperature increases during MD$_{\rm NA}$ until reheating completes at $a_R$.}
	\label{fig:n0k3}	
\end{figure}

We analyze the case of $k-n+5/2>0$, by checking how each term in the radiation energy density scales,
\begin{equation}
\label{eq:rhoR_anlytic}
\rho_R 
= \left(X_{R,i}^{(4-n)/4} + \beta\frac{X_{M}}{H_i} \frac{4-n}{4 \left(k-n+\frac{5}{2} \right)} \left(\left(\frac{a}{a_i}\right)^{k-n+\frac{5}{2}}-1\right)\right)^{\scalebox{1.01}{$\frac{4}{4-n}$}}\left(\frac{a}{a_i}\right)^{-4}.
\end{equation}
The terms involving $X_{R,i}$ and $-1$ evolve adiabatically, i.e.~scaling as $(a/a_i)^{-4}$, while the remaining $k$-dependent term scales with the exponent
\begin{equation}
\label{eq:scaling_k}
\left(k-n+\frac{5}{2} \right)\times \frac{4}{4-n}-4 =\frac{4k-6}{4-n}.
\end{equation}
If $k-n+5/2>0$ is satisfied, then this exponent is larger than $-4$ and the $k$-dependent term will eventually become the dominant source of radiation. This set of solutions lies above the gray boundary in Fig.~\ref{fig:phase-diagram}. When a stricter condition $k>3/2$ is also met as in the white region, the scale factor's exponent given by Eq.~(\ref{eq:scaling_k}) is positive, which immediately implies that the temperature \emph{increases} with the scale factor. Since we are interested in the period where the radiation sourced from matter dominates over the initial radiation energy density, the solution is simplified to
\begin{equation}
X_R = \Gamma \frac{X_{M,i}}{H_i}\left(\frac{a}{a_i}\right)^{-k+n} \frac{4-n}{4 \left(k-n+\frac{5}{2} \right)} \left(\left(\frac{a}{a_i}\right)^{k-n+\frac{5}{2}}-1\right).
\end{equation}
In terms of the energy densities, $\rho_M$ and $\rho_R$, the solution reads
\begin{equation}
\label{eq:rho_R}
\rho_R = \rho_M \frac{\Gamma}{H}\frac{4-n}{4 \left(k-n+\frac{5}{2} \right)} \left(1-\left(\frac{a_i}{a}\right)^{k-n+\frac{5}{2}}\right).
\end{equation}
This solution is consistent with our earlier analysis in Eq.~(\ref{eq:rhoR_T_est}), where $k>3/2$ implies an era of increasing temperature. On the other hand, $k-n+5/2<0$ implies adiabatic evolution because the initial radiation remains the dominant component based on Eq.~(\ref{eq:rhoR_anlytic}). An example of a cosmological history with this solution is shown in Fig.~\ref{fig:n3k3} for the values of $n=3$ and $k=3$ and another is shown in Fig.~\ref{fig:n0k3} for the values of $n=0$ and $k=3$. Both of these cases belong to the white region of Fig.~\ref{fig:phase-diagram}. In both Figs.~\ref{fig:n3k3} and \ref{fig:n0k3}, the upper panels show these solutions in an inflationary reheating scenario and assume that the dissipation products dominate immediately after inflation. The lower panels show these solutions in a more general context where reheating follows after an early RD era.

\begin{figure}
	\includegraphics[width= \linewidth]{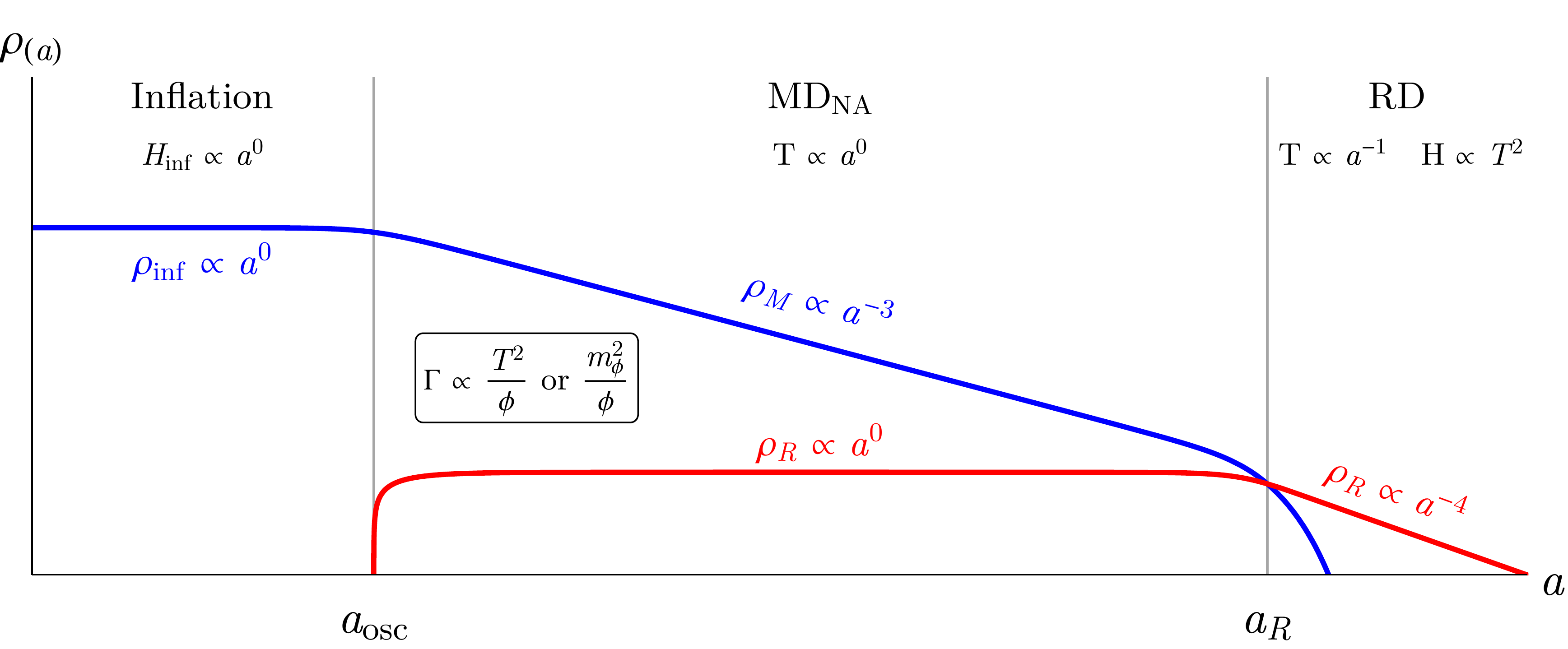}
	\includegraphics[width= \linewidth]{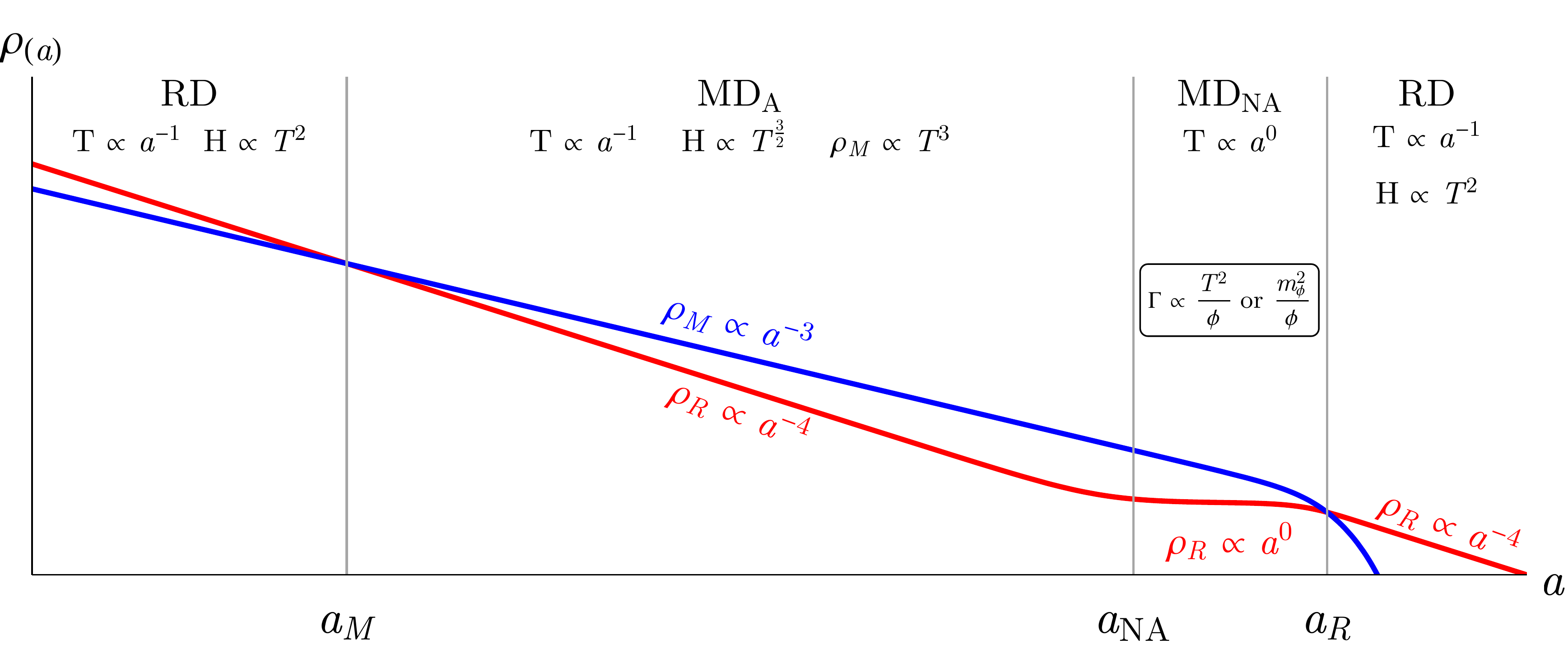}
	\caption{Evolution of the matter energy density $\rho_M$ and radiation energy density $\rho_R$ as a function of the scale factor $a$ with dissipation parameters $n=2$, $k=3/2$ and $n=0$, $k=3/2$. This can come from dissipation of a coherently oscillating scalar field $\phi$ with a rate $\Gamma\propto T^{2}/\phi$ and $\phi^{-1}\propto a^{3/2}$ (assuming a quadratic potential) as shown in Sec.~\ref{sec:models}. In the upper panel, a radiation bath is created shortly after inflation from perturbative decays of matter $\rho_M$. In this case, reheating is preceded by a period of constant temperature. The lower panel is a numerical solution in a generic reheating scenario. The dissipation rate causes a period of constant temperature from the start of a matter dominated non-adiabatic era at $a_{\rm NA}$ until reheating at $a_R$. During the period of constant temperature, dissipation replenishes the radiation as quickly as it redshifts. This is an inflection point of decreasing and increasing temperature of the diagram in Fig.~\ref{fig:phase-diagram} with $k=3/2$.}
	\label{fig:n2k3/2}	
\end{figure}

On the black line in Fig.~\ref{fig:phase-diagram}, $k=3/2$ and the temperature reaches a constant value, which can be computed from Eq.~(\ref{eq:rho_R}), during the matter dominated era until reheating ends. It is depicted in Fig.~\ref{fig:n2k3/2} that the constant temperature is attained when the radiation produced from dissipation dominates the initial radiation bath, either right after inflation as in the upper panel or after the initial bath redshifts to become negligible as in the lower panel. 

In all of the cases above, both dissipation and reheating complete when the dissipation rate and Hubble parameter are comparable $\Gamma \simeq H$.
If instead $k-n+5/2<0$, the $k$-dependent term will become irrelevant at late times, at which point $X_R$ becomes a constant. In this case, some amount of matter is transferred into radiation before the interaction rate is inefficient and dissipation does not complete. After $X_R$ approaches to the asymptotic value, radiation evolves adiabatically. This branch of solutions is in the overlap of the blue and gray regions of Fig.~\ref{fig:phase-diagram}, i.e.~between the black dotted line and the gray boundary. 

\subsection{Solution for $n>4$}
\label{subsec:large_n}

For $n>4$, although the solution rewritten here as
\begin{equation}
\label{eq:XR_n_ge_4}
X_R = \left(X_{R,i}^{(4-n)/4} - \beta \frac{X_{M}}{H_i} \frac{4-n}{4 \left(k-n+\frac{5}{2} \right)} \left(1-\left(\frac{a}{a_i}\right)^{k-n+\frac{5}{2}}\right) \right)^{\scalebox{1.01}{$\frac{4}{4-n}$}},
\end{equation}
is mathematically identical to Eq.~(\ref{eq:full_expression}), the following analysis is significantly different. When $n>4$, the exponent $4/(4-n)$ is negative. Unlike the case with $n<4$, a small $X_{R,i}$ term is not negligible and instead important for understanding the qualitative behavior. 

\begin{figure}
	\includegraphics[width= \linewidth]{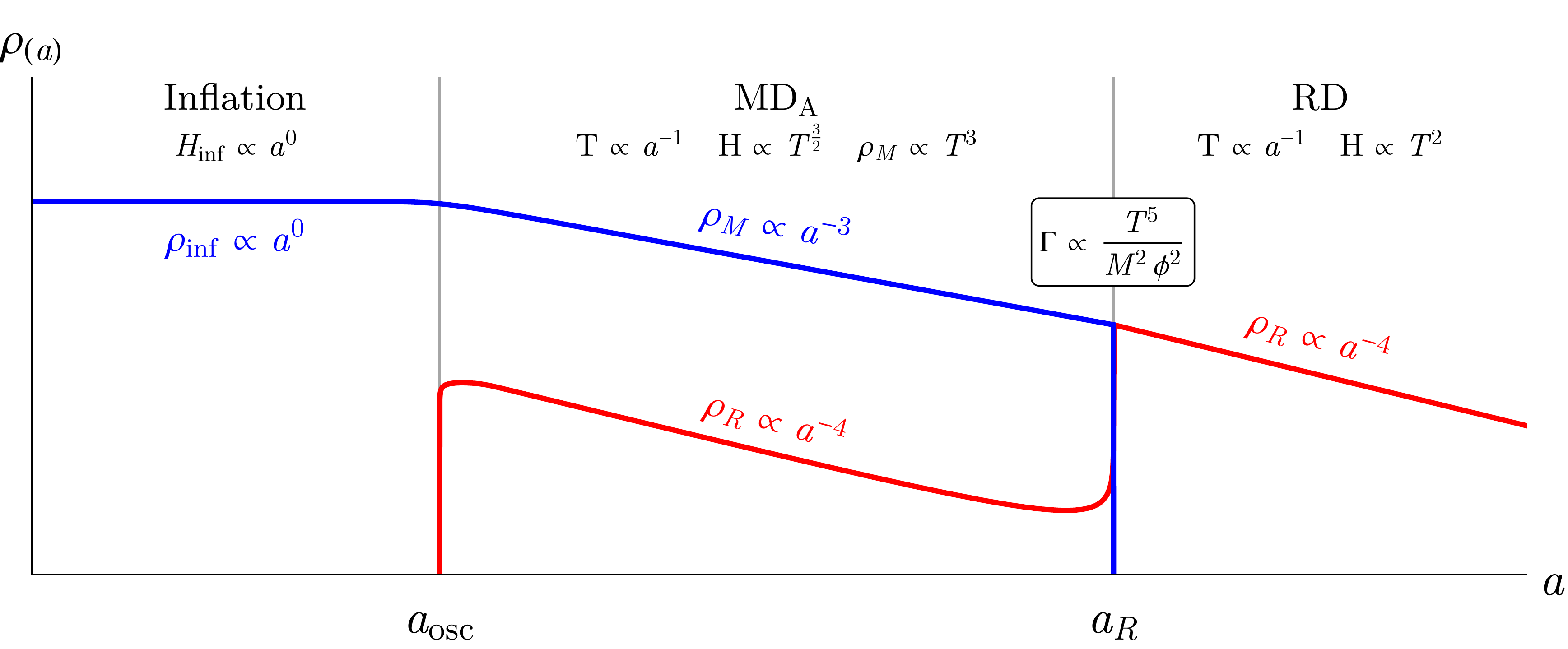}
	\includegraphics[width= \linewidth]{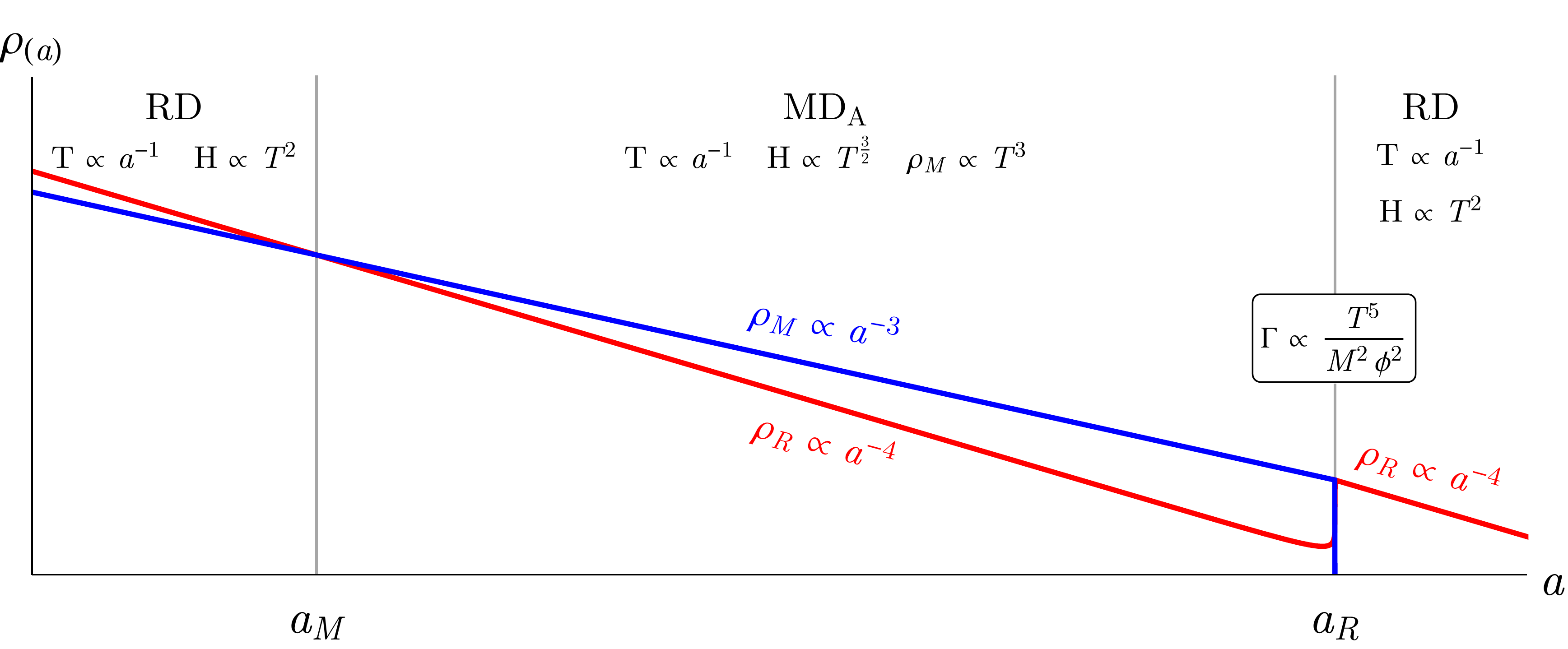}
	\caption{Evolution of the matter energy density $\rho_M$ and radiation energy density $\rho_R$ as a function of the scale factor $a$ with dissipation parameters $n=5$ and $k=3$. The evolution where the dissipation rate arises from a higher dimensional operator involving a rotating or fluctuating scalar field $\phi\propto a^{-3/2}$ with a quadratic potential as discussed in Sec.~\ref{sec:models}. In the upper panel, perturbative decays populate the initial radiation energy density at $a_{\rm osc}$. The radiation created from dissipation is initially subdominant to the radiation from decays, but rapid completion of dissipation due to non-linear effects takes hold moments before $a_R$, giving $T_R = T_{\rm max}$. In the lower panel, we show a numerical solution of the same behavior in a general reheating scenario, where an early radiation dominated era transitions to an early matter dominated era at $a_M$. Once again, the initial radiation dominates until moments before reheating at $a_R$, when non-linear effects rapidly complete dissipation. In both panels, the MD eras are mostly in the adiabatic phase except when the temperature increases abruptly.}
	\label{fig:n5k3}	
\end{figure}

We now discuss the set of solutions inside the part of the orange hatched region that is outside the gray region of Fig.~\ref{fig:phase-diagram}. For $k-n+5/2>0$, the $k$-dependent term will dominate at late times and the entire $\beta$-dependent term becomes increasingly negative and will eventually cancel $X_{R,i}^{(4-n)/4}$ at a finite $a$, which is much larger than $a_i$ unless $\beta$ is large. The current case $n>4$ implies that the entire expression in Eq.~(\ref{eq:XR_n_ge_4}) is raised to a negative power and a cancellation between the two terms in the expression causes a sharp increase in $X_R$ and thus the temperature. At this point, $X_R$ becomes sufficiently large in a short time so that $X_M$ is entirely depleted. Therefore, dissipation always completes for any nonzero value of $\beta$. Since the power law dependence of $\rho_R$ on $a$ given in Eq.~(\ref{eq:scaling_k}) is negative in this case, the temperature continues to decrease until the moment before reheating completes. The final sudden increase in the temperature causes the rate to overshoot the threshold of efficient dissipation, i.e.~$\Gamma \gg H$, depleting matter instantaneously. The reheat temperature can be estimated from conservation of energy $\rho_M (T_R) = \rho_R (T_R)$, where $\rho_M (T_R)$ is obtained from the adiabatic evolution up until the aforementioned cancellation occurs. This behavior is non-linear and best visualized in Fig.~\ref{fig:n5k3}. This phenomenon is physically understood as follows. Efficient dissipation increases the temperature, which in turn enhances the dissipation rate sharply due to the $T^n$ dependence. This feedback enhancement occurs during a non-adiabatic phase for all $n > 0$; nevertheless, the behavior of a sudden completion of dissipation is apparent for $n \geq 4$ because a rate highly sensitive to the temperature, i.e.~large $n$, is required to exhibit this phenomenon. We note that for inflationary reheating $T_{\rm max} = T_R$ is possible as long as $T_R$ is larger than the initial temperature right after inflation, which is true unless $\beta$ is too small.

\begin{figure}
	\includegraphics[width= \linewidth]{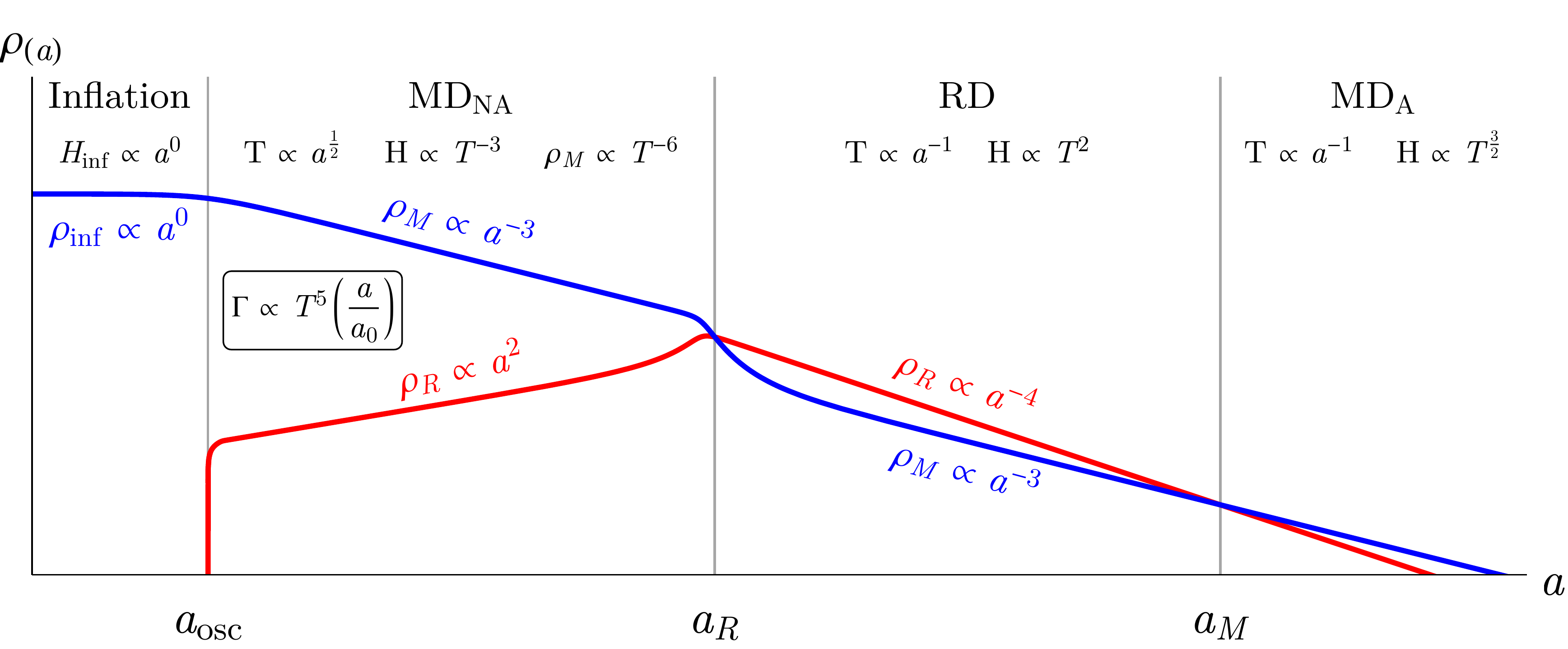}
	\caption{Evolution of the energy densities $\rho_M$ and $\rho_R$ where $\Gamma\propto T^5 a$  so $n = 5$ and $k = 1$. For these values of $n$ and $k$, dissipation will not complete and tuning is required for a prolonged, but incomplete, period of dissipation. In this cosmology, the temperature will increase at a rate $T\propto a^{1/2}$ as $\rho_R\propto a^2$ and the maximum temperature corresponds to the reheat temperature. Still, matter is not completely depleted and the radiation dominated era after this incomplete dissipation is followed by a second matter dominated era.}
	\label{fig:n5k1}	
\end{figure}

For $k-n+5/2<0$, the $k$-dependent term will vanish at late times and the entire $\beta$-dependent term approaches a constant value
\begin{equation}
Z \equiv - \beta \frac{X_{M}}{H_i} \frac{4-n}{4 \left(k-n+\frac{5}{2} \right)} < 0.
\end{equation}
If $|Z| \ll X_{R,i}^{(4-n)/4}$, then the final $X_R$ is only slightly modified from $X_{R,i}$, implying that reheating does not occur. If $|Z| \gg X_{R,i}^{(4-n)/4}$, then $X_{R,i}^{(4-n)/4} + Z$ will exhibit a cancellation at a value of $a$ only slightly larger than $a_i$. The evolution is qualitatively the same as instantaneous reheating. If $|Z| \simeq X_{R,i}^{(4-n)/4}$ instead, a similar situation occurs except that $X_M$ is only partially depleted and dissipation does not complete. An example of this type is illustrated in Fig.~\ref{fig:n5k1} with a full numerical solution. A rigorous numerical analysis confirms that dissipation never completes unless $\beta$ is sufficiently large and in this case one reproduces the usual instantaneous reheating. An extended period of reheating is only possible with a tuned $\beta$. The temperature monotonically increases with a tuned $\beta$ during reheating when the power of $\rho_R \propto a^{\frac{6-4k}{n-4}}$ based on Eq.~(\ref{eq:scaling_k}) is positive, which is the case for $k < 3/2$. This case falls into the purple hatched region of Fig.~\ref{fig:phase-diagram}. For $k > 3/2$, the temperature increases only in the last stage of reheating when the aforementioned cancellation is occurring. This corresponds to the overlap of the gray region and orange hatched region of Fig.~\ref{fig:phase-diagram}. We note that the discussion in this paragraph is only applicable to inflationary reheating because otherwise dissipation would have been completed in early times $a \ll a_i$ for generic reheating.

\subsection{Solution for $n=4$}
A special solution of Eq.~(\ref{eq:full_expression}) exists in the case of $n=4$. Here, the radiation energy density takes the solution in the form of an exponential,
\begin{equation}
X_R = X_{R,i} \exp\left[\beta\ \frac{X_{M}}{H_i} \frac{2}{2k-3} \left(\left(\frac{a}{a_i}\right)^{k-\frac{3}{2}}-1\right)\right],
\end{equation}
\begin{equation}
\rho_R = \rho_{R,i}\left(\frac{a}{a_i}\right)^{-4} \exp\left[\Gamma \frac{\rho_{M}}{\rho_R H} \frac{1}{k-\frac{3}{2}} \left(1-\left(\frac{a_i}{a}\right)^{k-\frac{3}{2}}\right)\right].
\end{equation}
This solution ultimately displays similar behavior as the adjacent $n>4$ regions. In particular, Fig.~\ref{fig:phase-diagram} is segmented into a black dotted line and a black dashed line at $n=4$. On the dashed line where $k>3/2$, there can be a period of increasing temperature as in the region with $n>4$ and $k>3/2$. On the black dotted line where $k<3/2$, the temperature decreases throughout the evolution and reheating only occurs when parameters are tuned, as in the region for $n>4$ and $k<3/2$.

\section{Reheating Scenarios Arising from Particle Physics Models}
\label{sec:models}

We now discuss the physical origins that give rise to the dependence of the dissipation rate on the temperature $T$ and the scale factor $a$ as parameterized by $\Gamma \propto T^n a^k$ in Eq.~(\ref{eq:Gamma_n_k}). 

The simplest scenario is the perturbative decay so that the rate is a constant so $n = 0$ and $k = 0$. The evolution is illustrated in Fig.~\ref{fig:pert_decay}. Contrary to what the term reheating suggests, the temperature constantly drops shortly after the onset of reheating.

If the dissipation interaction results from an operator with dimension $4+p$ and the suppression scale $\Lambda$ of the operator is time-independent, the rate is expected by dimensional analysis to be $\Gamma \propto T^{1+2p}/\Lambda^{2p}$ or $m^{1+2p}/\Lambda^{2p}$ depending on whether the temperature or another mass scale $m$ is relevant. If there is no additional scale factor dependence, we obtain $n = 1+2p$ and $k=0$. 

On the other hand, if the suppression scale is given by the field value of a scalar $\phi$ that evolves in its potential $V(\phi)$, then $k$ is nonzero due to the redshift of $\phi$. For example, the higher dimensional operator in the effective field theory can be a result of integrating out a heavy state $\psi$ whose mass $m_\psi(\phi) = y \phi$ is from the Yukawa interaction
\begin{align}
\mathcal{L} = y \phi \psi \bar\psi \, ,
\end{align}
where $y$ is the coupling constant.
When the low energy effective theory, obtained from integrating out $\psi\bar{\psi}$, contains a dimensionless coupling constant which logarithmically depends on the mass of $\psi\bar{\psi}$ through the renormalization, the interaction between $\phi$ and the thermal bath is suppressed by the field value of $\phi$. This is, for example, the case when $\psi\bar{\psi}$ is charged under a gauge symmetry and $\phi$ couples to the gauge bosons with the field strength $F_{\mu\nu}$,
\begin{align}
    {\cal L} \propto {\rm log}\frac{\phi}{T} F_{\mu\nu} F^{\mu\nu}.
\end{align}
The logarithmic dependence comes from the renormalization of the gauge coupling constant. The dissipation rate of $\phi$ is~\cite{Bodeker:2006ij, Laine:2010cq, Mukaida:2012qn}
\begin{align}
\label{eq:rate_5}
    \Gamma \propto \frac{T^3}{|\phi|^2},
\end{align}
which is valid when $T \gg m_\phi$. Otherwise, the rate is mass dependent instead
\begin{align}
\label{eq:rate_5_m}
    \Gamma \propto \frac{m_\phi^3}{|\phi|^2}.
\end{align}
When the low energy theory contains a dimension-five interaction which has a suppression scale $M$ and logarithmically depends on the mass of $\psi \bar{\psi}$ through the renormalization, the interaction of $\phi$ with the thermal bath is suppressed by $M$ and the field value $\phi$. One example is a dimension-five interaction between a scalar $S$ and a fermion $\eta$,
\begin{align}
\label{eq:Gamma_log_phi}
\mathcal{L} = \frac{1}{M} \log \frac{\phi}{T} \,  S S \eta \eta \,.
\end{align}
The suppression scale $M$ can be understood as originating from the mass $M_N$ of some heavy state $N$. When $N$ is integrated out, a Yukawa coupling $y_S S \eta N$ then gives the effective operator in Eq.~(\ref{eq:Gamma_log_phi}) with $1/M \simeq y_S^2 / M_N$. On the other hand, the logarithmic dependence on $\phi$ can originate from the renormalization of the mass $M_N$ and/or the Yukawa coupling $y_S$ by a coupling involving $\psi$ or $\bar{\psi}$.
The dissipation rate of $\phi$ is 
\begin{align}
\label{eq:rate_6}
\Gamma \propto 
\begin{dcases}
\frac{T^5}{M^2|\phi|^2} \ \ \ \ {\rm for} \ \ \ T \gg m_\phi \\
\frac{m_\phi^5}{M^2|\phi|^2} \ \ \ \ {\rm for} \ \ \ T \ll m_\phi 
\end{dcases}\, .
\end{align}

We consider the case where the field value of $\phi$ is initially large and decreases by redshifting during reheating towards $\phi =0$ where $\psi \bar{\psi}$ is massless. We also assume that the potential of $\phi$ is nearly quadratic. If $\phi$ is an inflaton, the large initial field value is realized in large field models. The Starobinsky model~\cite{Starobinsky:1980te} is consistent with the Cosmic Microwave Background observations~\cite{Akrami:2018odb}. A simple class of chaotic inflation~\cite{Linde:1983gd} is excluded by the upper bound on the tensor fraction~\cite{Akrami:2018odb}, but there exist models where the inflaton potential becomes flat at large field values so that the tensor fraction is suppressed~\cite{Destri:2007pv,Silverstein:2008sg,McAllister:2008hb,Takahashi:2010ky,Kallosh:2010ug,Harigaya:2012pg,Croon:2013ana,Nakayama:2013jka,Nakayama:2013txa,Li:2013nfa,Harigaya:2014qza,Harigaya:2014wta,Harigaya:2015pea,Harigaya:2017jny}. Alternatively, the curvature perturbations of the universe may be sourced by spectator field models such as a curvaton~\cite{Mollerach:1989hu,Linde:1996gt,Enqvist:2001zp,Lyth:2001nq,Moroi:2001ct,Enqvist:2002rf,Enqvist:2003mr,Harigaya:2019uhf} or modulated reheating~\cite{Dvali:2003em,Kofman:2003nx,Fujita:2016vfj,Karam:2020skk}. If $\phi$ is a generic scalar field, the large initial field value may be a result of a flat potential of $\phi$ or a negative mass term given by the coupling with the inflaton~\cite{Dine:1995uk}. For both the inflaton and the generic scalar field, the decrease of the field value towards the point where $\psi\bar{\psi}$ is massless is naturally explained if $\phi$ is charged under some (approximate) symmetry, such as a $Z_2$ or a $U(1)$ symmetry.

The field value $\phi$ may change rapidly, and one should use a dissipation rate averaged over a time period longer than the time scale of the change of $\phi$. In what follows, we discuss three possible scenarios for dynamics of $\phi$ in the potential: rotations, oscillations, and fluctuations.

We first discuss the dissipation rates given by Eqs.~(\ref{eq:rate_5}) and (\ref{eq:rate_5_m}). Possible dynamics for $\phi$ is the rotation of a complex field in the phase direction. Such rotations actually occur in Affleck-Dine baryogenesis~\cite{Affleck:1984fy,Dine:1995uk,Dine:1995kz}, baryogenesis from a complex inflaton field~\cite{Charng:2008ke,Hertzberg:2013jba,Hertzberg:2013mba} or from an axion field~\cite{Chiba:2003vp, Takahashi:2003db, Co:2019wyp,Co:2020xlh,Co:2020jtv},
magnetogenesis from an axion field~\cite{Kamada:2019uxp}, and the axion dark matter production with kinetic misalignment~\cite{Co:2019jts, Co:2020dya}. If the rotation is sufficiently circular so that the minimal value of $\lambda |\phi|$ during the cycle is above $T$ or $m_{\phi}$ for $T\gg m_\phi$ and $T\ll m_\phi$ respectively,
the averaged dissipation rate is given by
\begin{align}
\label{eq:Gamma_T3/phi2}
\Gamma_{\rm ave} \propto 
\begin{dcases}
\frac{T^3}{\bar{\phi}^2} \ \ \ \ {\rm for} \ \ \ T \gg m_\phi \\
\frac{m_\phi^3}{\bar{\phi}^2} \ \ \ \ {\rm for} \ \ \ T \ll m_\phi 
\end{dcases}\, .
\end{align}
Here $\bar{\phi}$ is the amplitude of the rotation and redshifts as $a^{-3/2}$ for a quadratic potential in the radial direction, giving $n = 3$ and $k = 3$ for $T \gg m_\phi$ and $n = 0$ and $k = 3$ for $T \ll m_\phi$. As derived in Sec.~\ref{subsec:small_n} and also summarized in Fig.~\ref{fig:phase-diagram}, these scenarios lead to increasing temperature during reheating. In Fig.~\ref{fig:n3k3} for $T \gg m_\phi$ and Fig.~\ref{fig:n0k3} for $T \ll m_\phi$, the upper (lower) panels result if the radiation energy density is initially dominated by the component created from dissipation (by the initial component).
When $T\ll m_\phi$ initially and later $T$ becomes larger than $m_\phi$, the scaling of the radiation energy changes accordingly. The same is also true for other cases described below.

Another possibility is that $\phi$ oscillates around the minimum of the potential. For the oscillation in a vacuum potential, the averaged dissipation rate is given by~\cite{ Mukaida:2012qn}
\begin{align}
\label{eq:Gamma_T2/phi}
\Gamma_{\rm ave} \propto 
\begin{dcases}
\frac{T^2}{\overline\phi} \ \ \ \ {\rm for} \ \ \ T \gg m_\phi \\
\frac{m_\phi^2}{\overline\phi} \ \ \ \ {\rm for} \ \ \ T \ll m_\phi 
\end{dcases}\, ,
\end{align}
where $\overline\phi$ is the amplitude of the oscillation. This scaling assumes that the non-perturbative process such as parametric resonance is ineffective.%
\footnote{Parametric resonant production of $\psi$ is ineffective if the adiabaticity condition $\dot m_\psi / m_\psi^2 < 1 $ is satisfied throughout the cycle with $m_\psi^2 \simeq \lambda^2 \phi^2 + T^2$. Given that $\dot m_\psi / m_\psi^2$ is maximized when $\lambda \phi \simeq T$, the adiabaticity condition is satisfied when $\lambda \bar{\phi} < T^2 / m$. Even if this condition is violated, since $\psi$ is a fermion, Pauli-blocking prevents the effective transfer of the energy of $\phi$ into $\psi$.}
Such a scaling for $T\gg m_\phi$ can be understood as follows.
The dissipation rate is enhanced when $m_\psi(\phi)$ gets small during the oscillation. However, the effective operator is valid only when $m_\psi(\phi) > T$ so that $\psi$ can be integrated out. Therefore, the estimate is $\Gamma_{\rm ave} \propto \delta \times T^3 / \phi_{\rm th}^2$ where $m_\psi(\phi_{\rm th}) \equiv T$ and $\delta$ is the fraction of the period when $\phi \sim \mathcal{O}({\phi_{\rm th}})$. Here $\delta$ is estimated by the ratio of the duration when $\phi$ is $\mathcal{O}({\phi_{\rm th}})$, i.e.~$\phi_{\rm th} / \dot\phi \simeq \phi_{\rm th} / m_\phi \overline\phi$, to the period of the cycle, $m_\phi^{-1}$. Now with $\delta = \phi_{\rm th} / \overline\phi$ and $\phi_{\rm th} = T/y$, one obtains the scaling in Eq.~(\ref{eq:Gamma_T2/phi}).
For $T\ll m_\phi$, one should replace $T$ with $m_{\phi}$ in the above discussion. For a quadratic potential with $\phi \propto a^{-3/2}$, Fig.~\ref{fig:n2k3/2} illustrates the evolution which exhibits an era with constant temperature as elaborated in Sec.~\ref{subsec:small_n}. An epoch with a constant temperature was first pointed out in Ref.~\cite{Co:2017pyf} in the context of Higgs dynamics with a large initial field value after inflation. The upper (lower) panel applies when the initial radiation energy density is dominated by the component created from dissipation (by the initial component). Here we have simply assumed a scaling of $\bar{\phi} \propto a^{-3/2}$ throughout the evolution, which is not a good approximation near the end of reheating. Instead, $\phi$ will decrease due to depletion of $\rho_M$, the rate is enhanced, and thus the temperature will increase towards the completion of reheating.

The last possibility is that $\phi$ is not in the form of a coherent condensate but large fluctuations. This scenario results from a non-perturbative process such as parametric resonance.
\newpage
\noindent
When the oscillating or rotating $\phi$ has a self-interaction, the mass of $\phi$ oscillates and the fluctuations of $\phi$ are amplified by parametric resonance. Once the amplitude of the fluctuations becomes as large as that of the coherent condensate, $\phi$ is better described as a fluctuating field rather than as a coherent condensate. Furthermore, the initial coherent condensate is destroyed and converted to fluctuating excitations due to the back-reaction. The fragmentation of the $\phi$ condensate into fluctuations can also occur by the parametric resonance production of an additional field $\chi$ if $\phi$ has a sufficient coupling with $\chi$. This arises in a variety of particle physics considerations, e.g.~preheating after inflation~\cite{Dolgov:1989us, Traschen:1990sw, Kofman:1994rk, Shtanov:1994ce, Kofman:1997yn} and the production of dark matter in the early Universe~\cite{Co:2017mop,Dror:2018pdh,Co:2020dya}. The resultant estimation of the averaged dissipation rate is expected to be similar to that of the rotation because, unlike the coherent oscillation case, the fluctuating field rarely evolves close to the origin and the corresponding enhancement is absent. We note that, in the case of $\chi$ production, back-reaction may not occur if $\chi$ is efficiently dissipated while being produced. Dissipation of $\phi$ may already complete by a very efficient parametric resonant production of $\chi$ and the dissipation of it  -- the scenario considered in instant preheating~\cite{Felder:1998vq}.

After discussing possible dynamics of $\phi$, we now return to the dissipation rate given by Eq.~(\ref{eq:rate_6}), where $n=5$. A rotating or fluctuating $\phi$ gives $k=3$, while an oscillating $\phi$ gives $k=3/2$. Fig.~\ref{fig:n5k3} illustrates the evolution for the rotating or fluctuating $\phi$, which exhibits an era with decreasing temperature followed by a sudden non-adiabatic phase near the end of the matter dominated era. We emphasize again that such an intriguing phenomenon is not a result of an approximation and can be understood from the analytic derivation in Sec.~\ref{subsec:large_n}.

For completeness, we show in Fig.~\ref{fig:n5k1} the cosmological evolution for $\Gamma \propto T^5 a$. As eleborated in Sec.~\ref{subsec:large_n}, the solution in this category, i.e.~the purple hatched region in Fig.~\ref{fig:phase-diagram}, requires fine tuning in the rate for a prolonged period of reheating. In this fine-tuned case, the temperature constantly increases during the matter dominated era. On the other hand, if the rate is larger (smaller) than the required tuned value, dissipation completes instantaneously after inflation (never completes). Unlike other cases discussed in this paper, a physical origin of such a dissipation rate is currently lacking even though the prolonged period of reheating with monotonically increasing temperature is interesting.

\section{Discussions}
\label{sec:discussions}
In our efforts to attain a more comprehensive understanding of reheating, we have found new dynamics of reheating in the early universe. These new possibilities are studied during an early matter dominated era with a generic dissipation rate dependent on the temperature $T$ and scale factor $a$ as in Eq.~(\ref{eq:Gamma_n_k}). The rate is parameterized as $\Gamma \propto T^n a^k$. In contrast to the usual perturbative decay scenario ($n = 0$ and $k = 0$) where the universe cools during reheating at a rate slower than that with adiabatic expansion, we show that it is possible for the temperature to remain constant or increase during reheating. Our understanding of the $n$-$k$ parameter space is summarized in Fig.~\ref{fig:phase-diagram}. In particular, for values $n<4$ and $k>3/2$ or $n\geq 4$ and $k<3/2$ the temperature increases monotonically throughout reheating. Furthermore, in the case where $n\geq4$ and $k>3/2$, the temperature will increase abruptly only during the final stage of the matter dominated era. If $k=3/2$ we find a period of constant temperature. Among these categorizations, we also find that the matter energy density can only be completely depleted outside the gray region, i.e.~$k - n + 5/2 > 0$.

Not only are the scenarios presented mathematically possible but we also motivate regions of the $n$-$k$ parameter space with different field theoretical origins of the dissipation processes. Our analysis in Sec.~\ref{sec:models} shows that the dynamics of a scalar field $\phi$ leads to a non-trivial dissipation rate. If the scalar field is complex and rotates in its potential or obtains large fluctuations because of parametric resonance, we show that dissipation can proceed via a rate $\Gamma \propto T^3/\phi^2$ or $\Gamma \propto m_\phi^3/\phi^2$ with $k=3$ for a quadratic potential of $\phi$. These models are in the region of $n<4$ and $k>3/2$ of our $n$-$k$ parameter space and the temperature will increase throughout reheating.
Furthermore, coherent oscillations around the origin of a scalar field can lead to a dissipation rate $\Gamma \propto T^2 /\bar{\phi}$ with oscillation amplitude $\bar{\phi} \propto a^{-3/2}$ for a quadratic potential. We relate such a scaling to $n<4$ and $k=3/2$ where the temperature remains constant until the end of reheating. We also show an example where rotations or fluctuations of a scalar field can lead to a rate $\Gamma \propto T^5/\phi^2$, which implies $n=5$ and $k=3$ for a quadratic potential. Our analysis shows that models with $n\geq4$ and $k>3/2$ have the interesting quality of sudden non-linear dissipation at the end of reheating. Compelling models of inflation, baryogenesis, the solution to the strong CP problem, and dark matter production contain scalar fields which rotate or oscillate in their potentials. In addition to the original motivation, these models introduce interesting cosmological phenomenology.

The importance of the trait $T_R = T_{\rm max}$, a direct consequence of constant or increasing temperature, cannot be overstated as it can resolve existing complications in models with high reheat temperatures and additional spontaneously broken symmetries. Even if the phase transition temperature is above $T_R$, in conventional reheating where $T_{\rm max} > T_R$, the symmetries may be thermally restored after inflation. If subsequent symmetry breaking produces stable topological defects, they may cause cosmological problems. Symmetry restoration can be prevented when $T_{\rm max}$ is only as large as $T_R$. Another phenomenological feature of models with increasing temperature is that phase transitions can occur twice, once as the temperature increases beyond the critical temperature during reheating, and again after reheating as the temperature decreases with adiabatic expansion. If the phase transition is of first order, the additional first order phase transition may generate observable gravitational waves. Lastly, if dark matter is produced before or during this novel cosmological era, the new temperature evolution must be considered. The thermal history of the universe affects thermal relic abundances and can even impact dark matter production mechanisms as scaling changes before $T_R$. For physics dependent on $T_{\rm max}$, relating $T_{\rm max}$ to features of the theory instead of inflationary initial conditions improves the predictability of the theory.

We now contrast instantaneous reheating with the cosmological evolution for a constant or increasing temperature. Although they both share the feature of $T_{\rm max} = T_R$, the existence of a non-inflating universe before $T = T_{\rm max}$ in the latter case can lead to drastically different phenomenological consequences. The Hubble rate during inflation $H_{\rm inf}$ is necessarily tied to $T_R$ for instantaneous reheating, while $H_{\rm inf}$ takes on a much larger value for non-instantaneous reheating.
For example, for $T_{\rm max} \sim 10^9$ GeV, instantaneous reheating requires a very small Hubble rate during inflation, $H_{\rm inf}\sim 1$ GeV. A phase transition may occur twice since the temperature can cross the same value twice in both scenarios. However, for instantaneous reheating, the first phase transition can occur only at the completion of the reheating. In addition, the first phase transition may not be thermal since the reheating occurs quickly and the thermal equilibrium may not be established at that point.

In summary, by studying the structure of the Boltzmann equations with a generic dissipation rate dependent on the temperature and scale factor, we categorize novel cosmological eras. We demonstrate theoretical origins of dissipation rates responsible for most of the categorized cosmological eras. These origins are not simply plausible but may play a role in theoretically motivated scenarios and lead to physically observable consequences. 

\vspace{0.5cm}

 {\bf Acknowledgment.}---%
The authors thank Edward Kolb for discussions, Andrew Long for comments on the manuscript, and Robert Brandenberger for the insightful remarks on inflationary reheating. The work was supported in part by the DoE Early Career Grant DE-SC0019225 (R.C.), the DoE grant DE-SC0009988 (K.H.), and the Raymond and Beverly Sackler Foundation Fund (K.H.).

\bibliography{Incr_temp}

\end{document}